\documentclass[manyauthors,nocleardouble,COMPASS]{cernphprep}
\usepackage{amssymb}
\usepackage{amsmath}
\usepackage{times}
\usepackage{epsfig}
\usepackage{colordvi}
\usepackage{graphicx}
\usepackage{wrapfig,rotating}
\usepackage{hyperref} 
\usepackage{caption}
\usepackage{multicol}
\usepackage{booktabs}
\usepackage{multirow}
\usepackage{lineno}

\newcommand{\AUT} {$A^{\sin(\phi-\phi_S)}_{\text{UT}}$}
\newcommand{\AUTphiS} {$A^{\sin\phi_S}_{\text{UT}}$}
\newcommand{\ALTphiS} {$A^{\cos\phi_S}_{\text{LT}}$}
\newcommand{\AUTphi} {$A^{\sin(2\phi-\phi_S)}_{\text{UT}}$}
\newcommand{\re}{\mathrm{Re}\,}
\newcommand{\im}{\mathrm{Im}\,}

\makeatletter
\newsavebox\myboxA
\newsavebox\myboxB
\newlength\mylenA

\newcommand*\xoverline[2][0.6]{%
    \sbox{\myboxA}{$\m@th#2$}%
    \setbox\myboxB\null
    \ht\myboxB=\ht\myboxA%
    \dp\myboxB=\dp\myboxA%
    \wd\myboxB=#1\wd\myboxA
    \sbox\myboxB{$\m@th\overline{\copy\myboxB}$}
    \setlength\mylenA{\the\wd\myboxA}
    \addtolength\mylenA{-\the\wd\myboxB}%
    \ifdim\wd\myboxB<\wd\myboxA%
       \rlap{\hskip 0.7\mylenA\usebox\myboxB}{\usebox\myboxA}%
    \else
        \hskip -0.7\mylenA\rlap{\usebox\myboxA}{\hskip 0.5\mylenA\usebox\myboxB}%
    \fi}
\makeatother
\DeclareSymbolFont{letters}     {OML}{cmm}{m}{it}
\DeclareSymbolFont{symbols}     {OMS}{cmsy}{m}{n}
\DeclareSymbolFont{largesymbols}{OMX}{cmex}{m}{n}
\begin{document}
\begin{titlepage}
\PHnumber{2013--191}
\PHdate{3 October 2013}

\title{Transverse target spin asymmetries in exclusive $\rho^0$ muoproduction}

\Collaboration{The COMPASS collaboration}
\ShortAuthor{The COMPASS collaboration}

\begin{abstract}
Exclusive production of $\rho^0$ mesons was studied at the COMPASS experiment by
scattering 160~GeV/$c$ muons off transversely polarised protons. Five
single-spin and three double-spin azimuthal asymmetries were measured as a
function of $Q^2$, $x_{Bj}$, or $p_{T}^{2}$.  The $\sin \phi_S$ asymmetry is
found to be $-0.019 \pm 0.008(stat.) \pm 0.003(syst.)$. All other asymmetries
are also found to be of small magnitude and consistent with zero within
experimental uncertainties.  Very recent calculations using a GPD-based model
agree well with the present results. The data is interpreted as evidence for the
existence of chiral-odd, transverse generalized parton distributions.
\end{abstract}
\vfill
\Submitted{(submitted to Phys.~Lett.~B)}
\end{titlepage}

{\pagestyle{empty}
%
%

\section*{The COMPASS Collaboration}
\label{app:collab}
\renewcommand\labelenumi{\textsuperscript{\theenumi}~}
\renewcommand\theenumi{\arabic{enumi}}
\begin{flushleft}
C.~Adolph\Irefn{erlangen},
M.G.~Alekseev\Irefn{triest_i},
V.Yu.~Alexakhin\Irefn{dubna},
Yu.~Alexandrov\Irefn{moscowlpi}\Deceased,
G.D.~Alexeev\Irefn{dubna},
A.~Amoroso\Irefn{turin_u},
V.~Andrieux\Irefn{saclay},
A.~Austregesilo\Irefnn{cern}{munichtu},
B.~Bade{\l}ek\Irefn{warsawu},
F.~Balestra\Irefn{turin_u}
% ...
\end{flushleft}

%
%

\begin{Authlist}
\item \Idef{bielefeld}{Universit\"at Bielefeld, Fakult\"at f\"ur Physik, 33501 Bielefeld, Germany\Arefs{f}}
\item \Idef{bochum}{Universit\"at Bochum, Institut f\"ur Experimentalphysik, 44780 Bochum, Germany\Arefs{f}\Arefs{ll}}
\item \Idef{bonniskp}{Universit\"at Bonn, Helmholtz-Institut f\"ur  Strahlen- und Kernphysik, 53115 Bonn, Germany\Arefs{f}}
\item \Idef{bonnpi}{Universit\"at Bonn, Physikalisches Institut, 53115 Bonn, Germany\Arefs{f}}
\item \Idef{brno}{Institute of Scientific Instruments, AS CR, 61264 Brno, Czech Republic\Arefs{g}}
\item \Idef{calcutta}{Matrivani Institute of Experimental Research \& Education, Calcutta-700 030, India\Arefs{h}}
\item \Idef{dubna}{Joint Institute for Nuclear Research, 141980 Dubna, Moscow region, Russia\Arefs{i}}
\item \Idef{erlangen}{Universit\"at Erlangen--N\"urnberg, Physikalisches Institut, 91054 Erlangen, Germany\Arefs{f}}
\item \Idef{freiburg}{Universit\"at Freiburg, Physikalisches Institut, 79104 Freiburg, Germany\Arefs{f}\Arefs{ll}}
\item \Idef{cern}{CERN, 1211 Geneva 23, Switzerland}
\item \Idef{liberec}{Technical University in Liberec, 46117 Liberec, Czech Republic\Arefs{g}}
\item \Idef{lisbon}{LIP, 1000-149 Lisbon, Portugal\Arefs{j}}
\item \Idef{mainz}{Universit\"at Mainz, Institut f\"ur Kernphysik, 55099 Mainz, Germany\Arefs{f}}
\item \Idef{miyazaki}{University of Miyazaki, Miyazaki 889-2192, Japan\Arefs{k}}
\item \Idef{moscowlpi}{Lebedev Physical Institute, 119991 Moscow, Russia}
\item \Idef{munichlmu}{Ludwig-Maximilians-Universit\"at M\"unchen, Department f\"ur Physik, 80799 Munich, Germany\Arefs{f}\Arefs{l}}
\item \Idef{munichtu}{Technische Universit\"at M\"unchen, Physik Department, 85748 Garching, Germany\Arefs{f}\Arefs{l}}
\item \Idef{nagoya}{Nagoya University, 464 Nagoya, Japan\Arefs{k}}
\item \Idef{praguecu}{Charles University in Prague, Faculty of Mathematics and Physics, 18000 Prague, Czech Republic\Arefs{g}}
\item \Idef{praguectu}{Czech Technical University in Prague, 16636 Prague, Czech Republic\Arefs{g}}
\item \Idef{protvino}{State Research Center of the Russian Federation, Institute for High Energy Physics, 142281 Protvino, Russia}
\item \Idef{saclay}{CEA IRFU/SPhN Saclay, 91191 Gif-sur-Yvette, France\Arefs{ll}}
\item \Idef{telaviv}{Tel Aviv University, School of Physics and Astronomy, 69978 Tel Aviv, Israel\Arefs{m}}
\item \Idef{triest_i}{Trieste Section of INFN, 34127 Trieste, Italy}
\item \Idef{triest}{University of Trieste, Department of Physics and Trieste Section of INFN, 34127 Trieste, Italy}
\item \Idef{triestictp}{Abdus Salam ICTP and Trieste Section of INFN, 34127 Trieste, Italy}
\item \Idef{turin_u}{University of Turin, Department of Physics and Torino Section of INFN, 10125 Turin, Italy}
\item \Idef{turin_i}{Torino Section of INFN, 10125 Turin, Italy}
\item \Idef{turin_p}{University of Eastern Piedmont, 15100 Alessandria,  and Torino Section of INFN, 10125 Turin, Italy}
\item \Idef{warsaw}{National Centre for Nuclear Research, 00-681 Warsaw, Poland\Arefs{n} }
\item \Idef{warsawu}{University of Warsaw, Faculty of Physics, 00-681 Warsaw, Poland\Arefs{n} }
\item \Idef{warsawtu}{Warsaw University of Technology, Institute of Radioelectronics, 00-665 Warsaw, Poland\Arefs{n} }
\item \Idef{yamagata}{Yamagata University, Yamagata, 992-8510 Japan\Arefs{k} }
\end{Authlist}
%
%
\vspace*{-\baselineskip}\renewcommand\theenumi{\alph{enumi}}
\begin{Authlist}
\item \Adef{a}{Also at IST, Universidade T\'ecnica de Lisboa, Lisbon, Portugal}
\item \Adef{aa}{Also at Department of Physics, Pusan National University, Busan 609-735, Republic of Korea and at Physics Department, Brookhaven National Laboratory, Upton, NY 11973, U.S.A. }
\item \Adef{bb}{Supported by the DFG Research Training Group Programme 1102  ``Physics at Hadron Accelerators''}
\item \Adef{b}{Also at Chubu University, Kasugai, Aichi, 487-8501 Japan\Arefs{k}}
\item \Adef{c}{Also at KEK, 1-1 Oho, Tsukuba, Ibaraki, 305-0801 Japan}
\item \Adef{cc}{Also at Moscow Institute of Physics and Technology, Moscow Region, 141700, Russia}
\item \Adef{y}{present address: National Science Foundation, 4201 Wilson Boulevard, Arlington, VA 22230, United States}
\item \Adef{x}{present address: RWTH Aachen University, III. Physikalisches Institut, 52056 Aachen, Germany}
\item \Adef{e}{Also at GSI mbH, Planckstr.\ 1, D-64291 Darmstadt, Germany}
\item \Adef{f}{Supported by the German Bundesministerium f\"ur Bildung und Forschung}
\item \Adef{g}{Supported by Czech Republic MEYS Grants ME492 and LA242}
\item \Adef{h}{Supported by SAIL (CSR), Govt.\ of India}
\item \Adef{i}{Supported by CERN-RFBR Grants 08-02-91009 and 12-02-91500}
\item \Adef{j}{\raggedright Supported by the Portuguese FCT - Funda\c{c}\~{a}o para a Ci\^{e}ncia e Tecnologia, COMPETE and QREN, Grants CERN/FP/109323/2009, CERN/FP/116376/2010 and CERN/FP/123600/2011}
\item \Adef{k}{Supported by the MEXT and the JSPS under the Grants No.18002006, No.20540299 and No.18540281; Daiko Foundation and Yamada Foundation}
\item \Adef{l}{Supported by the DFG cluster of excellence `Origin and Structure of the Universe' (www.universe-cluster.de)}
\item \Adef{ll}{Supported by EU FP7 (HadronPhysics3, Grant Agreement number 283286)}
\item \Adef{m}{Supported by the Israel Science Foundation, founded by the Israel Academy of Sciences and Humanities}
\item \Adef{n}{Supported by the Polish NCN Grant DEC-2011/01/M/ST2/02350}
\item [{\makebox[2mm][l]{\textsuperscript{*}}}] Deceased
\end{Authlist}

\clearpage}
\section{Introduction}

\label{sect:intro}
The spin structure of the nucleon is a key issue in experimental and theoretical
research since a few decades. The most general information on the partonic
structure of hadrons is contained in the generalised parton correlation
functions (GPCFs)~\cite{Meissner:2009ww,Lorce:2013pza}, which parameterise the
fully unintegrated, off-diagonal parton-parton correlators for a given
hadron. These GPCFs are 'mother distributions' of the generalised parton
distributions (GPDs) and the transverse momentum dependent parton distributions
(TMDs), which can be considered as different projections or limiting cases of
GPCFs. While GPDs appear in the QCD-description of hard exclusive processes such
as deeply virtual Compton scattering (DVCS) and hard exclusive meson production
(HEMP), TMDs can be measured in reactions like semi-inclusive deep inelastic
scattering (SIDIS) or Drell-Yan processes. The GPDs and TMDs provide
complementary 3-dimensional pictures of the nucleon. In particular, when
Fourier-transformed to impact parameter space and for the case of vanishing
longitudinal momentum transfer, GPDs provide a three dimensional description of
the nucleon in a mixed momentum-coordinate space, also known as `nucleon
tomography'~\cite{Burkardt:2000za,Burkardt:2002hr}.  Moreover, GPDs and TMDs
contain information on the orbital motion of partons inside the nucleon.
 
The process amplitude for hard exclusive meson production by {\it longitudinal}
virtual photons was proven rigorously to factorise into a hard-scattering part
and a soft part~\cite{Radyushkin:1996ru,Collins:1996fb}. The hard part is
calculable in perturbative QCD (pQCD). The soft part contains GPDs to describe
the structure of the probed nucleon and a distribution amplitude (DA) to
describe the one of the produced meson. This collinear factorisation holds in
the generalised Bjorken limit of large photon virtuality $Q^2$ and large total
energy in the virtual-photon nucleon system, $W$, but fixed $x_{Bj}$, and for
$|t|/Q^2\ll1$. Here $t$ is the four-momentum transfer to the proton and $x_{Bj}=
Q^2/2M_{p}\nu$, where $\nu$ is the energy of the virtual photon in the lab frame
and $M_p$ the proton mass.

For hard exclusive meson production by {\it transverse} virtual photons, no
proof of collinear factorisation exists. In phenomenological pQCD-inspired
models $k_\perp$ factorisation is used, where $k_\perp$ denotes the parton
transverse momentum. In the model of
Refs.~\cite{Goloskokov:2005sd,Goloskokov:2007nt,Goloskokov:2008ib},
electroproduction of a light vector meson V at small $x_{Bj}$ is analysed in the
'handbag' approach, in which the amplitude of the process is a convolution of
GPDs with amplitudes for the partonic subprocesses $\gamma^{*} q \rightarrow V
q$ and $\gamma^{*} g \rightarrow V g$. Here, $q$ and $g$ denote quarks and
gluons, respectively. The partonic subprocess amplitudes, which comprise
corresponding hard scattering kernels and meson DAs, are calculated in the
modified perturbative approach where the transverse momenta of quark and
antiquark forming the vector meson are retained and Sudakov suppressions are
taken into account. The partons are still emitted and reabsorbed from the
nucleon collinear to the nucleon momentum. In such models, cross sections and
also spin-density matrix elements for HEMP by both longitudinal and transverse
virtual photons can be well described
simultaneously~\cite{Goloskokov:2005sd,Martin:1996bp}.

At leading twist, the chiral-even GPDs $H^f$ and $E^f$, where $f$ denotes a
quark of a given flavor or a gluon, are sufficient to describe exclusive vector
meson production on a spin 1/2 target. These GPDs are of special interest as
they are related to the total angular momentum carried by partons in the
nucleon~\cite{Ji:1996ek}. A variety of GPD fits using all existing DVCS proton
data has shown that the contributions of GPDs $H^f$ are dominant. They are
constrained
~\cite{Guidal:2004nd,Kumericki:2009uq,Goldstein:2010gu,Guidal:2013rya} over the
presently limited accessible $x_{Bj}$ range, by the very-low $x_{Bj}$ data of
the HERA collider and by the high $x_{Bj}$ data of HERMES and JLab. There exist
constraints on GPDs $E^f$ for valence quarks from fits to nucleon form factor
data~\cite{Diehl:2004cx}, HERMES transverse proton data~\cite{Airapetian:2008aa}
and JLab neutron data~\cite{Mazouz:2007aa}. A parameterisation of chiral-even
GPDs ~\cite{Goloskokov:2008ib}, which is consistent with the HEMP data of
HERMES~\cite{Airapetian:2009ad} and COMPASS~\cite{Adolph:2012ht}, was recently
demonstrated to successfully describe almost all existing DVCS
data~\cite{Kroll:2012sm}. This is clear evidence for the consistency of the
contemporary phenomenological GPD-based description of both DVCS and HEMP.

There exist also chiral-odd -- often called transverse -- GPDs, from which in
particular $H^f_T$ and $\xoverline{E}^f_T$ were shown to be
required~\cite{Goloskokov:2009ia,Goloskokov:2011rd} for the description of
exclusive $\pi^+$ electroproduction on a transversely polarised proton
target~\cite{Airapetian:2009ac}. It was recently shown \cite{PrivCommPK} that
the data analysed in this letter are also sensitive to these GPDs.

This Letter describes the measurement of exclusive $\rho^0$ muoproduction on
transversely polarised protons with the COMPASS apparatus. Size and kinematic
dependences of azimuthal modulations of the cross section with respect to beam
and target polarisation are determined and discussed, in particular in terms of
the above introduced chiral-odd GPDs.

\section{Formalism}
\label{sect:formalism}
\begin{figure}
 \begin{center}
\includegraphics[width=0.7\textwidth]{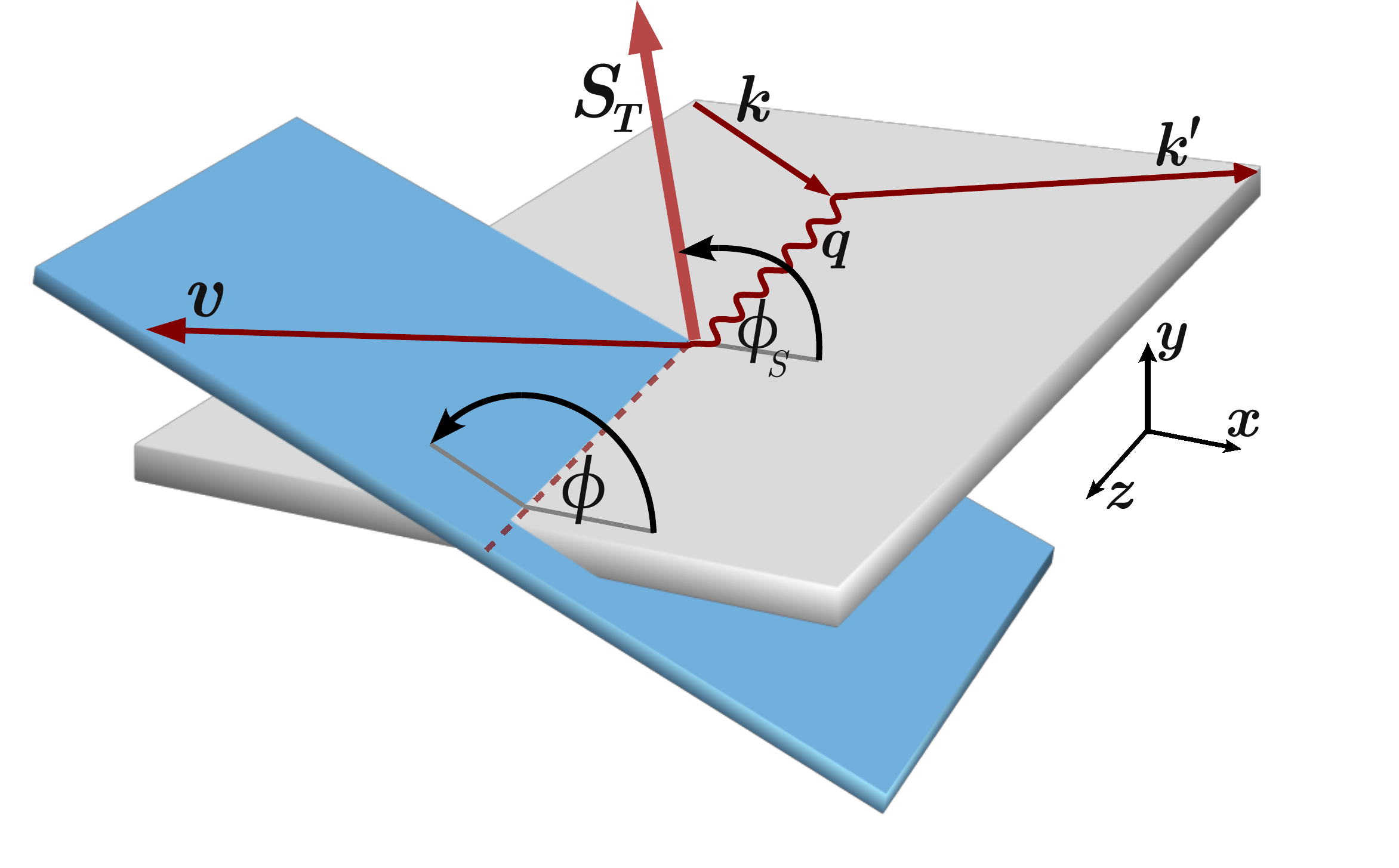}
\caption[Definition of the angles $\phi$ and $\phi_{s}$.]{Definition of the
angles $\phi$ and $\phi_{s}$. Here $\pmb{k}$, $\pmb{k'}$, $\pmb{q}$ and
$\pmb{v}$ represent three-momentum vectors of the incident and the scattered
muon, the virtual photon and the meson respectively. The symbol $S_T$ denotes
the component of the target spin vector perpendicular to the virtual-photon
direction.}
\label{fig:theory_angle}
 \end{center}
\end{figure}

The cross section for exclusive $\rho^0$ muoproduction, $\mu\,N \to
\mu'\,\rho^0\,N'$, on a transversely polarised target reads~\cite{Diehl:2005pc}:

\begin{align}
  \label{Xsection}
\frac{d\sigma}{dx_B\, dQ^2\, dt\, d\phi\, d\phi_S}
 &= \bigg[ \frac{\alpha_{\rm em}}{8\pi^3}\, 
\frac{y^2}{1-\varepsilon}\,
       \frac{1-x_{Bj}}{x_{Bj}}\, \frac{1}{Q^2} \bigg] \Bigg\{
\frac{1}{2} \Big( \sigma_{++}^{++} + \sigma_{++}^{--} \Big)
+ \varepsilon \sigma_{00}^{++} 
- \varepsilon \cos(2\phi)\, \re \sigma_{+-}^{++} \nonumber \\
& \hspace{1em} {}
- \sqrt{\varepsilon (1+\varepsilon)}\,
  \cos\phi\, \re (\sigma_{+0}^{++} + \sigma_{+0}^{--})
- P_\ell\, \sqrt{\varepsilon (1-\varepsilon)}\, 
           \sin\phi\, \im (\sigma_{+0}^{++} + \sigma_{+0}^{--})
\phantom{\Bigg[ \Bigg] }
\nonumber \\
&- S_T\, \bigg[
  \sin(\phi-\phi_S)\,
  \im (\sigma_{++}^{+-} + \varepsilon \sigma_{00}^{+-})
+ \frac{\varepsilon}{2} \sin(\phi+\phi_S)\, \im \sigma_{+-}^{+-}
\nonumber \\
& \hspace{1em} {}
+ \frac{\varepsilon}{2} \sin(3\phi-\phi_S)\, \im \sigma_{+-}^{-+}
+ \sqrt{\varepsilon (1+\varepsilon)}\, 
  \sin\phi_S\, \im \sigma_{+0}^{+-}
\nonumber \\
& \hspace{1em} {}
+ \sqrt{\varepsilon (1+\varepsilon)}\, 
  \sin(2\phi-\phi_S)\,  \im \sigma_{+0}^{-+}
\bigg]
\nonumber \\
&+ S_T P_\ell\, \bigg[
  \sqrt{1-\varepsilon^2}\, \cos(\phi-\phi_S)\, \re \sigma_{++}^{+-}
- \sqrt{\varepsilon (1-\varepsilon)}\, 
  \cos\phi_S\, \re \sigma_{+0}^{+-}
\nonumber \\
& \hspace{1em} {}
- \sqrt{\varepsilon (1-\varepsilon)}\, 
  \cos(2\phi-\phi_S)\,  \re \sigma_{+0}^{-+} 
\bigg] \Bigg\} .
\end{align}
 Here, $S_T$ is the target spin component perpendicular to the direction of the
virtual photon. The beam polarisation is denoted by $P_\ell$.  The azimuthal
angle between the lepton scattering plane and the production plane spanned by
virtual photon and produced meson is denoted by $\phi$, whereas $\phi_S$ is the
azimuthal angle of the target spin vector about the virtual-photon direction
relative to the lepton scattering plane (see Fig.~\ref{fig:theory_angle}). The
$S_{T}$ dependent part of Eq.~(\ref{Xsection}) contains eight different
azimuthal modulations: five sine modulations for the case of an unpolarised beam
and three cosine modulations for the case of a longitudinally polarised beam.
Neglecting terms depending on $m_{\mu}^2/Q^2$, where $m_{\mu}$ denotes the mass
of the incoming lepton, the virtual-photon polarisation parameter $\varepsilon$
describes the ratio of longitudinal and transverse photon fluxes and is given
by:
\begin{eqnarray}
  \label{eps-def}
\varepsilon
  = \frac{1 - y - \frac{1}{4} y^2 \gamma^2}{
          1 - y + \frac{1}{2} y^2 + \frac{1}{4} y^2 \gamma^2} \, ,
\qquad
\gamma = \frac{2M_px_{Bj}}{Q} \, .
\end{eqnarray} 
The symbols $\sigma^{\nu \lambda}_{\mu \sigma}$ in Eq.~(\ref{Xsection}) stand
for polarised photoabsorption cross sections or interference terms, which are
given as products of helicity amplitudes $\mathcal{M}$:
\begin{equation}
  \label{sigma-mn-ij}
\sigma^{\nu \lambda}_{\mu \sigma} = 
  \sum
   \mathcal{M}^*_{\mu' \nu',\mu \nu}
         \mathcal{M}_{\mu' \nu',\sigma \lambda},
\end{equation}
where the sum runs over $\mu'=0,\pm1$ and $\nu'=\pm1/2$. The helicity amplitude
labels appear in the following order: vector meson ($\mu'$), final-state proton
($\nu'$), photon ($\mu$ or $\sigma$), initial-state proton ($\nu$ or
$\lambda$). For brevity, the helicities $-1$, $-1/2$, $0$, $1/2$, $1$ will be
labelled by only their signs or zero, omitting 1 or 1/2, respectively. Also the
dependence of $\sigma^{\nu \lambda}_{\mu \sigma}$ on kinematic variables is
omitted.

The amplitudes of those cross section modulations that depend on target
polarisation are obtained from Eq.~(\ref{Xsection}) as follows:
\begin{alignat}{5}
  \label{eq:asym_def}
 &A_{\text{UT}}^{\sin(\phi-\phi_s) } &=&   - \frac{\rm{Im}(\sigma_{++}^{+-} + \varepsilon \; \sigma_{00}^{+-})}
  {\sigma_0},
\qquad \qquad
&A&_{\text{LT}}^{\cos(\phi-\phi_S)}  &=&  \frac{\re \sigma_{++}^{+-}}{\sigma_0},
\nonumber \\
&A_{\text{UT}}^{\sin(\phi+\phi_s) }  &=& - \frac{\im \sigma_{+-}^{+-}}{\sigma_0},
\qquad \qquad
&A&_{\text{LT}}^{\cos(\phi_s) } &=& - \frac{\re \sigma_{+0}^{+-}}{\sigma_0},
\nonumber \\ 
&A_{\text{UT}}^{\sin(3\phi-\phi_s) } &=& - \frac{\im \sigma_{+-}^{-+}}{\sigma_0},
\qquad \qquad
&A&_{\text{LT}}^{\cos(2\phi-\phi_s) } &=& - \frac{ \re \sigma_{+0}^{-+}}{\sigma_0},
\nonumber \\
 &A_{\text{UT}}^{\sin(\phi_s) } &=&   - \frac{\im \sigma_{+0}^{+-}}{\sigma_0},  \nonumber \\
 &A_{\text{UT}}^{\sin(2\phi-\phi_s) } &=&   - \frac{\im \sigma_{+0}^{-+}}{\sigma_0} .
\end{alignat}
Here, unpolarised (longitudinally polarised) beam is denoted by U (L) and
transverse target polarisation by T. The $\phi$-integrated cross section for
unpolarised beam and target, denoted by $\sigma_0$, is given as a sum of the
transverse and longitudinal cross sections:
\begin{eqnarray}
\label{eq:sigma0}
 \sigma_0 = \frac{1}{2} (\sigma_{++}^{++} + \sigma_{++}^{--}) + \varepsilon \sigma_{00}^{++} \,.
\end{eqnarray} 
The amplitudes given in Eq.~(\ref{eq:asym_def}) will be referred to as
asymmetries in the rest of the paper.

\section{Experimental set-up}
\label{sect:experiment}
The COMPASS experiment is situated at the high-intensity M2 muon beam of the
CERN SPS. A detailed description can be found in Ref.~\cite{Abbon:2007pq}.

The $\mu^+$ beam had a nominal momentum of 160 GeV/$c$ with a spread of 5\% and
a longitudinal polarisation of $P_{\ell}\approx-0.8$.  The data were taken at a
mean intensity of $3.5 \cdot 10^8\,\mu$/spill, for a spill length of about 10~s
every 40~s. A measurement of the trajectory and the momentum of each incoming
muon is performed upstream of the target.

The beam traverses a solid-state ammonia target that provides transversely
polarised protons. The target is situated within a large aperture magnet with a
dipole holding field of 0.5~T. The 2.5~T solenoidal field is only used when
polarising the target material. A mixture of liquid $^3$He and $^4$He is used to
cool the target to 50~mK. Ten NMR coils surrounding the target allow for a
measurement of the target polarisation $P_{T}$, which typical amounts to 0.8
with an uncertainty of 3\%. The ammonia is contained in three cylindrical target
cells with a diameter of 4~cm, placed one after another along the beam. The
central cell is 60~cm long and the two outer ones are 30~cm long, with 5 cm
space between cells. The spin directions in neighbouring cells are
opposite. Such a target configuration allows for a simultaneous measurement of
azimuthal asymmetries for the two target spin directions in order to become
independent of beam flux measurements. Systematic effects due to acceptance are
reduced by reversing the spin directions on a weekly basis. With the three-cell
configuration, the average acceptance for cells with opposite spin direction is
approximately the same, which leads to a further reduction of systematic
uncertainties.

The dilution factor $f$, which is the cross-section-weighted fraction of
polarisable material, is calculated for incoherent exclusive $\rho^0$ production
using the measured material composition and the nuclear dependence of the cross
section. It amounts typically to 0.25 \cite{Adolph:2012ht}.

The spectrometer consists of two stages in order to reconstruct scattered muons
and produced hadrons over wide momentum and angular ranges.  Each stage has a
dipole magnet with tracking detectors before and after the magnet, hadron and
electromagnetic calorimeters and muon identification. Identification of charged
tracks with a RICH detector in the first stage is not used in the present
analysis.

Inclusive and calorimetric triggers are used to activate data recording.
Inclusive triggers select scattered muons using pairs of hodoscopes and muon
absorbers whereas the calorimetric trigger relies on the energy deposit of
hadrons in one of the calorimeters. Veto counters upstream of the target are
used to suppress beam halo muons.

\section{Event selection and background estimation}
\label{sect:analysis}
The presented work is a continuation of the analysis of \AUT\ for exclusive
$\rho^0$ mesons produced off transversely polarised protons at COMPASS and it is
based on the same proton event sample as in Ref.~\cite{Adolph:2012ht}. The
essential steps of event selection and asymmetry extraction are summarized in
the following. The considered events are characterized by an incoming and a
scattered muon and two oppositely charged hadrons, $h^+h^-$, with all four
tracks associated to a common vertex in the polarised target. In order to select
events in the deep inelastic scattering regime and suppress radiative
corrections, the following cuts are used: $Q^2 > $ 1 (GeV/$c$)$^2$, 0.003
$<x_{Bj}<$ 0.35, $W>$ 5 GeV and 0.1$<y<$ 0.9, where $y$ is the fractional energy
of the virtual photon. The production of $\rho ^0$ mesons is selected in the
two-hadron invariant mass range 0.5 GeV/$c^2$ $< M_{\pi ^+\pi ^-} <$ 1.1
GeV/$c^2$, where for each hadron the pion mass hypothesis is assigned. This cut
is optimized towards high yield and purity of $\rho ^0$ production, as compared
to non-resonant $\pi^+\pi^-$ production.  The measurements are performed without
detection of the recoiling proton in the final state. Exclusive events are
selected by choosing a range in missing energy,
\begin{equation}
E_{\text{miss}} = \frac{(p+q-v)^2 - p^2}{2 M_p} = \frac{M_X^2 - M_p^2}{2  M_p},
\end{equation} 
where $M_X$ is the mass of the undetected recoiling system. This mass is
calculated from the four-momenta of proton, photon and meson, which are denoted
by $p$, $q$, and $v$ respectively. Although for exclusive events
$E_{\text{miss}}$ $\approx$ 0 holds, the finite experimental resolution is taken
into account by selecting events in the range $ |E_{\text{miss}} | <$ 2.5 GeV,
which corresponds to $0 \pm 2\sigma$ where $\sigma$ is the width of the Gaussian
signal peak. Non-exclusive background can be suppressed by cuts on the squared
transverse momentum of the vector meson with respect to the virtual photon
direction, $p_T^2 <$ 0.5 (GeV/$c$)$^2$, the energy of the $\rho ^0$ in the
laboratory system, $E_{\rho ^0} >$ 15 GeV, and the photon virtuality, $Q^2<$ 10
(GeV/$c$)$^2$. An additional cut $p_T^2>0.05$ (GeV/$c$)$^2$ is used to reduce
coherently produced events. As explained in Ref.~\cite{Adolph:2012ht} we use
$p_T^2$ rather than $t$. After the application of all cuts, the final data set
of incoherently produced exclusive $\rho ^0$ events consist of about 797000
events.  The average values of the kinematic variables are $\langle Q^2 \rangle
= 2.15 $ (GeV/$c$)$^2$, $\langle x_{Bj} \rangle = 0.039$, $\langle y \rangle =
0.24$, $\langle W \rangle = 8.13$ GeV, and $\langle p_T^2 \rangle = 0.18$
(GeV/$c$)$^2$.
\begin{figure}
\begin{center}
\includegraphics[trim=0mm 0mm 0mm 0mm, clip,scale=0.31]{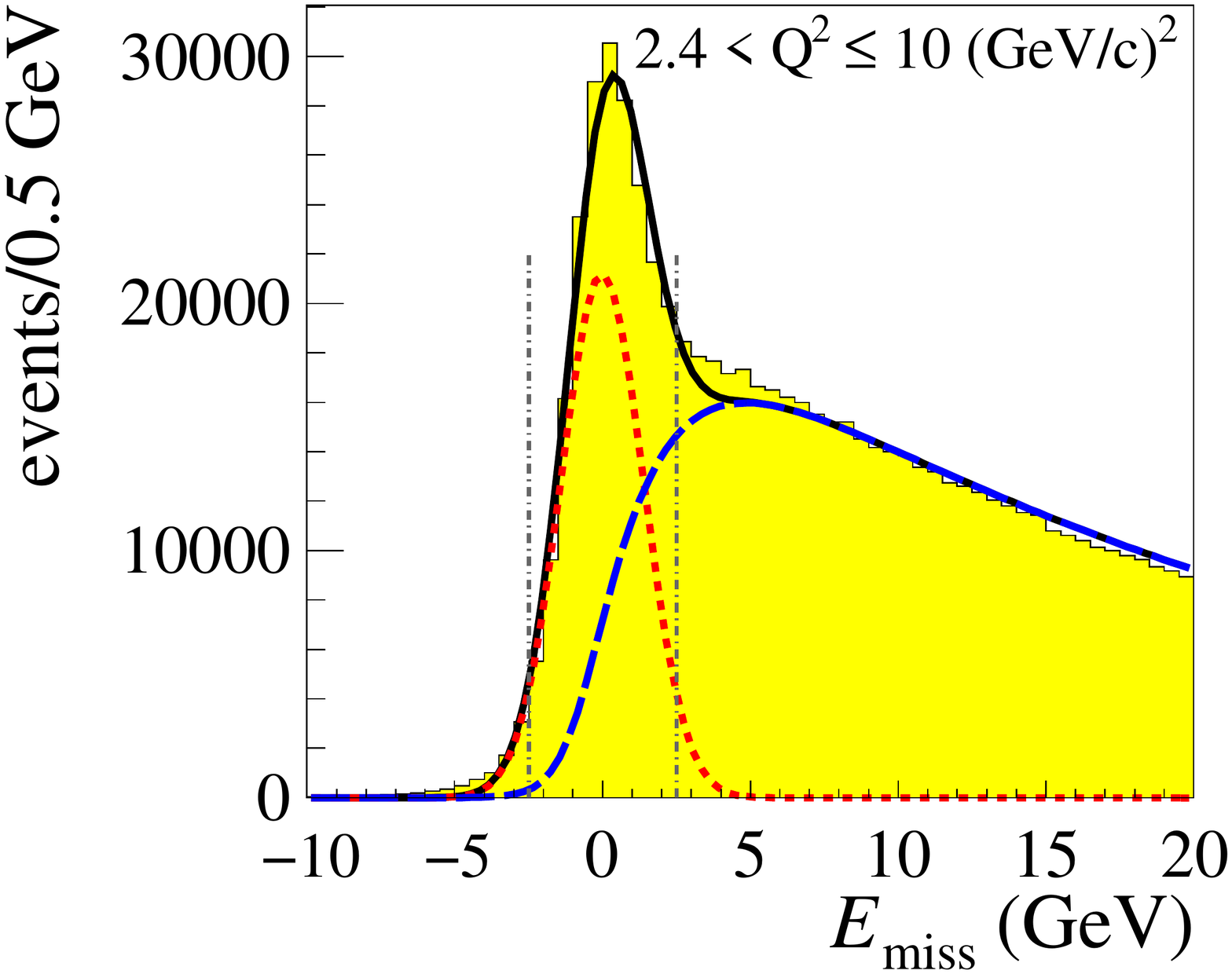}
\caption[]{The $E_{\text{miss}}$ distribution in the range 2.4 (GeV/$c$)$^2$ $<
Q^2 \le 10$ (GeV/$c$)$^2$, together with the signal plus background fits (solid
curve). The dotted and dashed curves represent the signal and background
contributions, respectively. In the signal region -2.5 GeV $< E_{\text{miss}} <
2.5$ GeV, indicated by vertical dash-dotted lines, the amount of semi-inclusive
background is 35\%.}
\label{fig:emiss}
\end{center}
\end{figure}
In order to correct for the remaining semi-inclusive background in the signal
region, the $E_{\text{miss}}$ shape of the background is parameterised for each
individual target cell in every kinematic bin of $Q^2$, $x_{Bj}$, or $p_{T}^2$
using a LEPTO Monte Carlo (MC) sample generated with COMPASS
tuning~\cite{Adolph:2012vj} of the JETSET parameters. The $h^+h^-$ MC event
sample is weighted in every $E_{\text{miss}}$ bin $i$ by the ratio of numbers of
$h^{\pm}h^{\pm}$ events from data and MC,
\begin{equation}
w_i=\frac{N^{h^+h^+}_{\text{i,data}}(E_{\text{miss}})+N^{h^-h^-}_{\text{i,data}}(E_{\text{miss}})}{N^{h^+h^+}_{\text{i,MC}}(E_{\text{miss}})+N^{h^-h^-}_{\text{i,MC}}(E_{\text{miss}})}, 
\end{equation}
which improves the agreement between data and MC
significantly~\cite{Adolph:2012ht}.

For each kinematic bin, target cell, and spin orientation a signal plus
background fit is performed, whereby a Gaussian function is used for the signal
shape, and the background shape is fixed by MC as described above. The fraction
of semi-inclusive background in the signal range is 22\%, nevertheless the
fraction strongly depends on kinematics and varies between 7\% and 40\%. An
example is presented in Fig.~\ref{fig:emiss}.  The background corrected
distributions, $N^{\text{sig}}_k(\phi,\phi_S)$, are obtained from the measured
distributions in the signal region, $N^{\text{sig,raw}}_k(\phi,\phi_S)$, and in
the background region 7 GeV $< E_{\text{miss}} <$ 20 GeV,
$N^{\text{back}}_k(\phi,\phi_S)$.  The distributions
$N^{\text{back}}_k(\phi,\phi_S)$ are rescaled with the estimated numbers of
background events in the signal region and afterwards subtracted from the
$N^{\text{sig,raw}}_k(\phi,\phi_S)$ distributions.

After the described subtraction of semi-inclusive background, the final sample
still contains diffractive events where the recoiling nucleon is in an excited
N$^{*}$ or $\Delta$ state (14\%), coherently produced $\rho ^0$ mesons ($\sim$
5\%), and non-resonant $\pi ^+ \pi ^-$ pairs ($<$ 2\%) \cite{Adolph:2012ht}. We
do not apply corrections for these contributions.

\section{Results and discussion}
\label{sect:results}
The asymmetries are evaluated using the background-corrected distributions
$N^{\text{sig}}_k(\phi,\phi_S)$ by combining data-taking periods with opposite
target polarisations. The events of the two outer target cells are summed
up. The number of exclusive $\rho ^0$ mesons as a function of $\phi$ and $\phi
_S$, where the index $j$ denotes the ($\phi$, $\phi _S$) bin, can be written for
every target cell $n$ as:
\begin{equation}
\label{eq:number_events}
N^{\pm} _{j,n} (\phi, \phi_S) = a^{\pm}_{j,n} \left( 1 \pm  A(\phi,\phi_S) \right).
\end{equation}
Here, $a^{\pm}_{j,n}$ is the product of spin-averaged cross section, muon flux,
number of target nucleons, acceptance, and efficiency of the spectrometer. The
angular dependence reads:
\begin{eqnarray}
A(\phi, \phi_S) &=& A_{\text{UT,raw}}^{\sin(\phi - \phi _S)} \sin(\phi - \phi _S)+ A_{\text{UT,raw}}^{\sin(\phi+\phi _S)} \sin(\phi+\phi _S) \nonumber \\ 
&& + { } A_{\text{UT,raw}}^{\sin(3\phi -\phi _S)} \sin(3\phi - \phi _S) 
+ A_{\text{UT,raw}}^{\sin(2\phi - \phi _S)} \sin(2\phi - \phi _ S) \nonumber \\
&& + { } A_{\text{UT,raw}}^{\sin(\phi _S)} \sin(\phi _S) + A_{\text{LT,raw}}^{\cos(\phi - \phi _S)} \cos(\phi - \phi _S) \nonumber \\
&& + { } A_{\text{LT,raw}}^{\cos (\phi _S)} \cos(\phi _S) + A_{\text{LT,raw}}^{\cos(2\phi -\phi _S)} \cos(2\phi - \phi _S) .
\end{eqnarray} 
The symbol $A^m_{\text{UT(LT),raw}}$ denotes the amplitude for the angular
modulation $m$. After the subtraction of semi-inclusive background, the ``raw''
asymmetries $A^m_{\text{UT, raw}}$ and $A^m_{\text{LT,raw}}$ are extracted from
the final sample using a two-dimensional binned maximum likelihood fit in $\phi$
and $\phi_S$. They are used to obtain the transverse target asymmetries
$A^m_{\text{UT}(\text{LT})}$ defined in Eq.~(\ref{eq:asym_def}) as:
\begin{alignat}{3}\label{eq:raw_asym}
 &A^m_{\text{UT}} &=&  \frac{A^m_{\text{UT,raw}}}{\langle f \cdot |P_{T}| \cdot D^m (\epsilon)\rangle} {} \, , \nonumber \\
&A^m_{\text{LT}} &=&  \frac{A^m_{\text{LT,raw}}}{\langle f \cdot |P_{T}| \cdot P_{\ell} \cdot D^m (\epsilon) \rangle}{} \, .
\end{alignat} 
Here, $P_T$ is used, which in COMPASS kinematics is a good approximation to
$S_T$. The depolarisation factors are given by:
\begin{alignat}{4}
\label{eq:dnn}
&&&D^{\sin(\phi - \phi _S)} &=& 1, \nonumber \\
&D^{\sin(\phi + \phi _S)} &=& D^{\sin(3\phi - \phi _S)} &=& \frac{\varepsilon}{2} \approx \frac{1-y}{1+(1-y)^2}, \nonumber \\
&D^{\sin(\phi _S)} &=& D^{\sin(2\phi - \phi _S)} &=& \sqrt{\varepsilon(1+\varepsilon)} \approx \frac{(2-y) \sqrt{2(1-y)}}{1+(1-y)^2}, \nonumber \\
&&&D^{\cos(\phi -\phi _S)} &=&\sqrt{1-\varepsilon^2} \approx \frac{y(2-y)}{1+(1-y)^2}, \nonumber \\
&D^{\cos\phi _S}&=& D^{\cos(2\phi - \phi _S)} &=& \sqrt{\varepsilon(1-\varepsilon)} \approx \frac{y\sqrt{2(1-y)}}{1+(1-y)^2}.
\end{alignat}
In order to estimate the systematic uncertainty of the measurements, we take
into account the relative uncertainty of the target dilution factor (2\%), the
target polarisation (3\%), and the beam polarisation (5\%). Combined in
quadrature this gives an overall systematic normalisation uncertainty of 3.6\%
for the asymmetries $A^m_{\text{UT}}$ and 6.2\% for $A^m_{\text{LT}}$.
Additional systematic uncertainties are obtained from separate studies of i) a
possible bias of the applied estimator, ii) the stability of the asymmetries
over data-taking time, and iii) the robustness of the applied background
subtraction method and the correction by the depolarization factors from
Eq.~(\ref{eq:dnn}). A summary of systematic uncertainties for the average
asymmetries can be found in Table ~\ref{tab:sys_summary}. The total systematic
uncertainty is obtained as a quadratic sum of these three components. In
Eq.~(\ref{Xsection}), $S_T$ is defined with respect to the virtual-photon
momentum direction, while in the experiment transverse polarization $P_T$ is
defined relative to the beam direction. The transition from $S_T$ to $P_T$
introduces in the cross section~\cite{Diehl:2005pc} the angle $\theta$ between
the virtual photon and the beam direction, which is small at COMPASS
kinematics. Additionally, some of the $A_{\text{UT(LT)}}$ asymmetries get mixed
with $A_{\text{UL(LL)}}$ asymmetries that are suppressed by $\sin\theta$. The
influence of the $\theta$-related corrections was studied in detail and found to
be negligible for all analysed asymmetries.

\begin{figure}
\begin{center}
\includegraphics[trim=0mm 0mm 0mm 0mm, clip,scale=0.46]{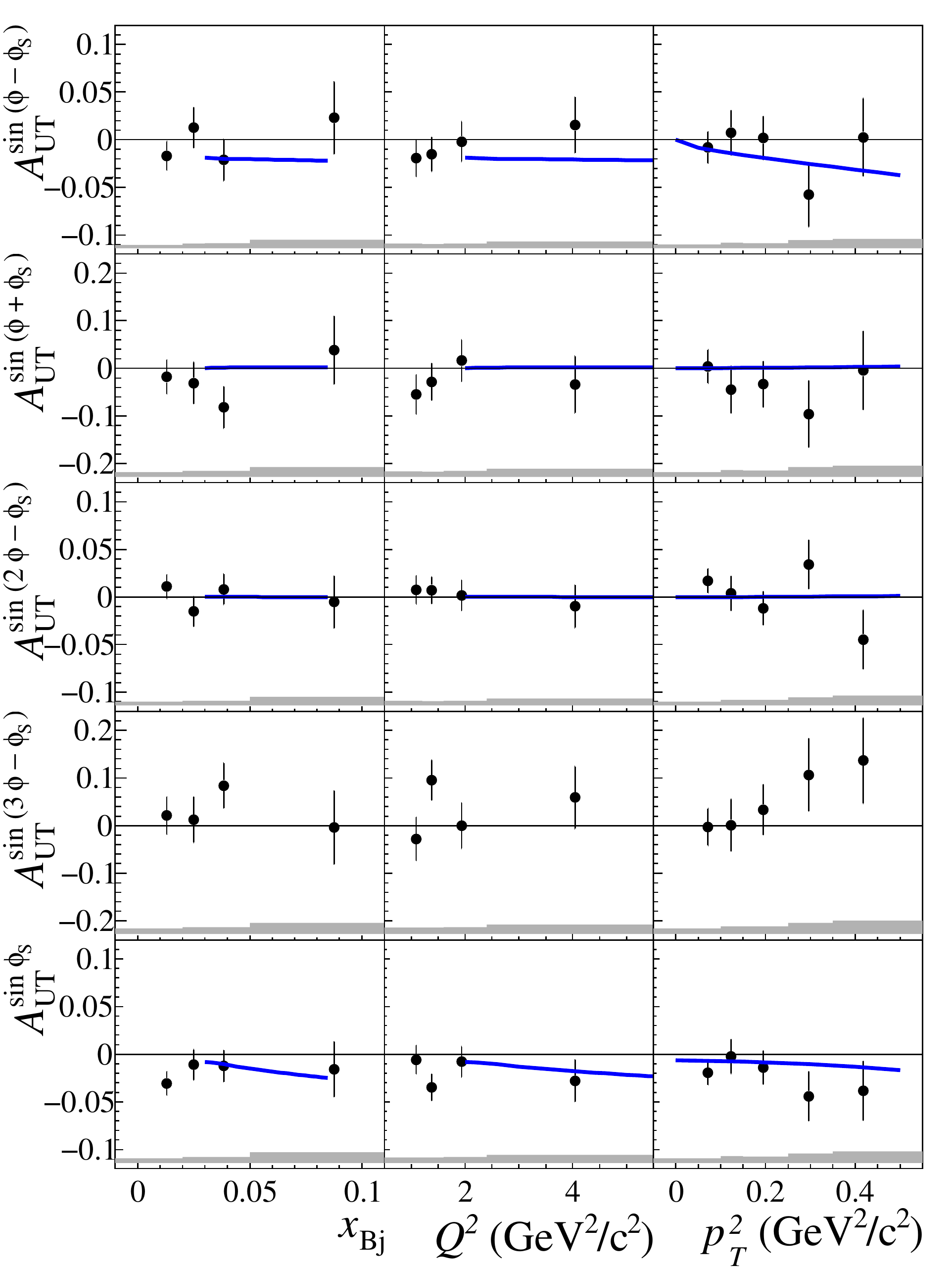}
\caption[]{Single-spin azimuthal asymmetries for a transversely (T) polarised
target and unpolarised (U) beam. The error bars (bands) represent the
statistical (systematic) uncertainties. The curves show the predictions of the
GPD model \cite{PrivCommPK}. They are calculated for the average $W$, $Q^2$ and
$p_T^2$ of our data set, $W = 8.1$ GeV/$c^2$ and $p^2_T$ = 0.2 (GeV/$c$)$^2$ for
the left and middle panels, and at $W = 8.1$ GeV/$c^2$ and $Q^2$ = 2.2
(GeV/$c$)$^2$ for the right panels.  The asymmetry
$A_{\text{UT}}^{\sin(3\phi-\phi_S)}$ is assumed to be zero in this model.}
\label{fig:aut}
\end{center}
\end{figure}

\begin{figure}
\begin{center}
\includegraphics[trim=0mm 0mm 0mm 0mm,clip,scale=0.46]{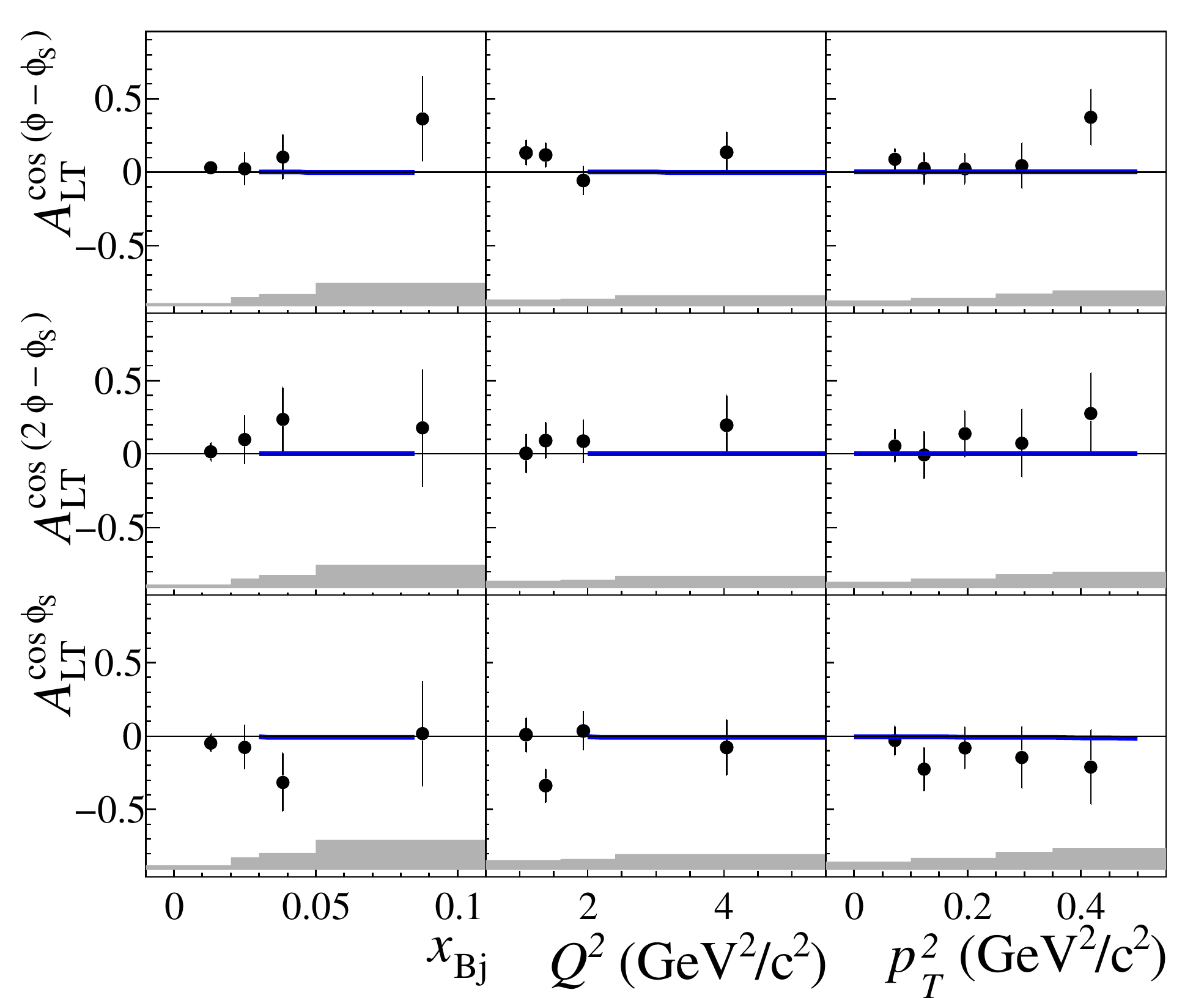}
\caption[]{Double-spin azimuthal asymmetries for a transversely (T) polarised
target and a longitudinally (L) polarised beam. The error bars (bands) represent
the statistical (systematic) uncertainties.  They are calculated for the average
$W$, $Q^2$ and $p_T^2$ of our data set, $W = 8.1$ GeV/$c^2$ and $p^2_T$ = 0.2
(GeV/$c$)$^2$ for the left and middle panels, and at $W = 8.1$ GeV/$c^2$ and
$Q^2$ = 2.2 (GeV/$c$)$^2$ for the right panels.
}
\label{fig:alt}
\end{center}
\end{figure}

\begin{figure}
\begin{center}
\includegraphics[trim=0mm 0mm 0mm 0mm, clip,scale=0.34]{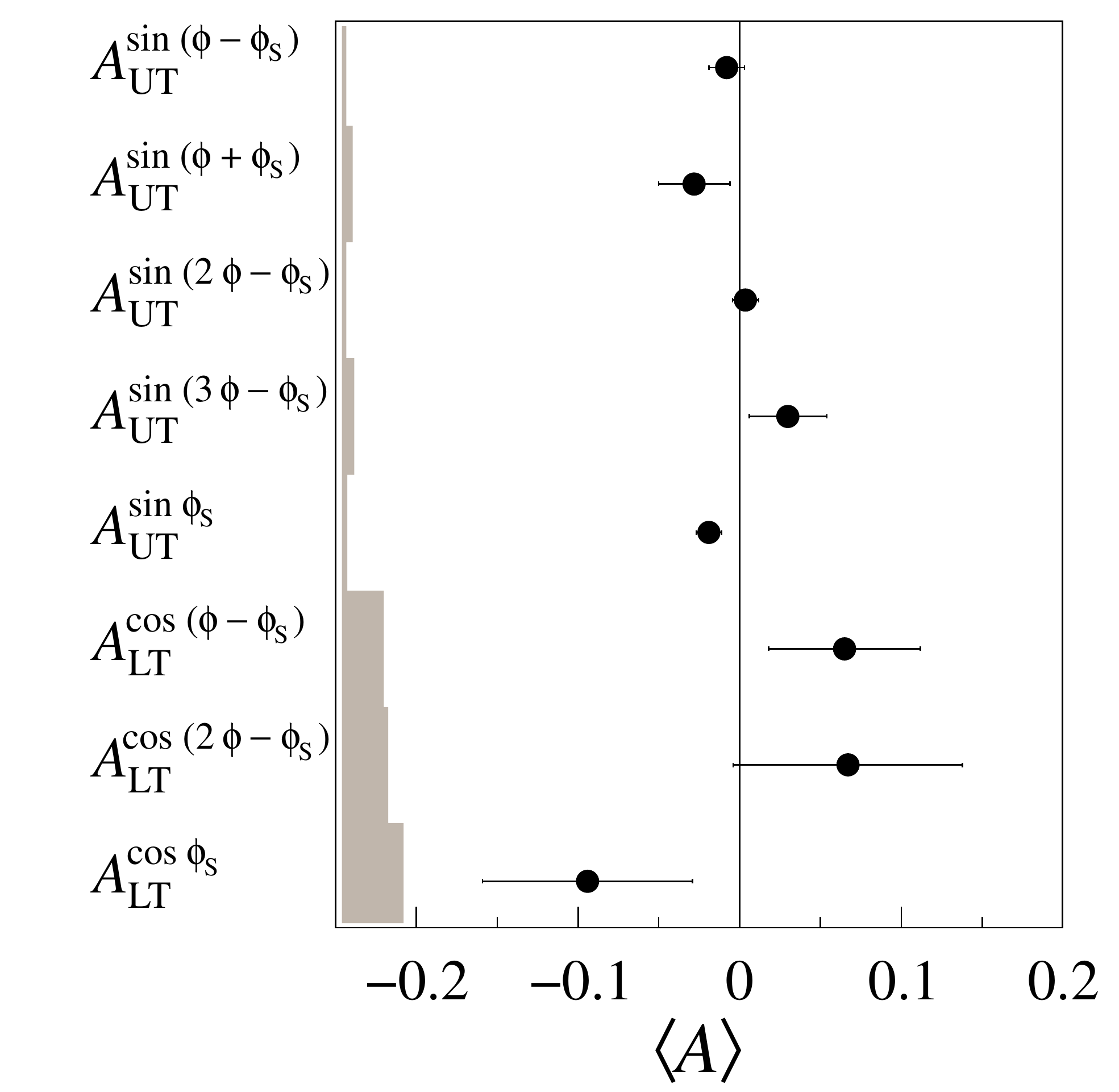}
\caption[]{Mean value $\langle A \rangle$ and the statistical error for every
modulation. The error bars (left bands) represent the statistical (systematic)
uncertainties.}
\label{fig:asym_mean}
\end{center}
\end{figure}

\begin{table}[!hf]
\begin{center}
\caption[]{Systematic uncertainties for the average asymmetries obtained from
the studies explained in the text.}
\begin{tabular}{|p{2.5cm} p{1.cm} p{1.cm} p{1.cm} | p{2.5cm} p{1.cm} p{1.cm} p{1.cm}| }
\hline
 & i) & ii) & iii) &  & i) & ii) & iii) \\
\hline
A$_{\text{UT}}^{\sin(\phi - \phi _S)}$ & 0.002 & 0.002 & 0.001  & A$_{\text{LT}}^{\cos(\phi - \phi _S)}$ & 0.005 & 0.011 & 0.023\\
A$_{\text{UT}}^{\sin(\phi+\phi _S)}$ & 0.004 & 0.004 & 0.004 & A$_{\text{LT}}^{\cos(2\phi -\phi _S)} $ & 0.016 & 0.016 & 0.018\\
A$_{\text{UT}}^{\sin(2\phi -\phi _S)}$ & 0.002 & 0.001 & 0.002 & A$_{\text{LT}}^{\cos (\phi _S)}$& 0.006 & 0.029 & 0.023\\
A$_{\text{UT}}^{\sin(3\phi - \phi _S)}$ & 0.006  & 0.003 & 0.003 & & & &\\
A$_{\text{UT}}^{\sin(\phi _S)}$ & 0.001  & 0.003 & 0.000 & & & &\\
\hline
\end{tabular}
\label{tab:sys_summary}
\end{center}
\end{table}
The results for the five single-spin and three double-spin asymmetries as a
function of $x_{Bj}$, $Q^2$, or $p_T^2$ are shown in Figs.~\ref{fig:aut} and
\ref{fig:alt}, respectively. Error bars show statistical uncertainties. The
systematic uncertainties are represented by grey shaded bands.  Average
asymmetry values for all modulations are given in Fig.~\ref{fig:asym_mean} and
Table~\ref{tab:asym_summary}. For three of them, the experimental precision is
as high as $\cal{O}$ ($\pm$ 0.01). All average asymmetry values are found to be
of small magnitude, below 0.1. Except \AUTphiS, all other average asymmetry
values are consistent with zero within experimental uncertainties. All results
are available in the Durham data base.
\begin{table}[hf!]
\setlength{\extrarowheight}{8pt}
\begin{center}
\caption[]{Average asymmetries with statistical and systematic uncertainties for
all measured modulations.}
\label{tab:asym_summary}
\begin{tabular}{|lr|lr|}
\hline
$A_{\text{UT}}^{\sin(\phi - \phi _S)}$ & $-0.008  \pm 0.011 \pm 0.003$ & $A_{\text{LT}}^{\cos(\phi - \phi _S)}$ & $0.065 \pm 0.047 \pm 0.026$\\
$A_{\text{UT}}^{\sin(\phi+\phi _S)}$ &  $-0.028 \pm 0.022 \pm 0.006$ &  $A_{\text{LT}}^{\cos(2\phi -\phi _S)}$ & $0.067 \pm 0.071 \pm 0.029$\\
$A_{\text{UT}}^{\sin(2\phi -\phi _S)}$ &  $0.004 \pm 0.008 \pm 0.003$ & $A_{\text{LT}}^{\cos (\phi _S)}$ &   $-0.094 \pm 0.065 \pm 0.038$ \\
$A_{\text{UT}}^{\sin(3\phi - \phi _S)}$ &  $0.03 \pm 0.024 \pm 0.008$ & & \\
$A_{\text{UT}}^{\sin(\phi _S)}$  & $-0.019 \pm 0.008 \pm 0.003$ & &\\
\hline
\end{tabular}
\end{center}
\end{table}

As already mentioned above, there exists presently only the model of
Refs.~\cite{Goloskokov:2005sd,Goloskokov:2007nt,Goloskokov:2008ib} to describe
hard exclusive $\rho^0$ leptoproduction using GPDs. It is a phenomenological
`handbag' approach based on $k_\perp$ factorisation, which also includes twist-3
meson wave functions. Calculations for the full set of five $A_{\text{UT}}$ and
three $A_{\text{LT}}$ asymmetries were performed very
recently~\cite{PrivCommPK}. They are shown in Figs.~\ref{fig:aut},~\ref{fig:alt}
as curves together with the data points. Of particular interest is the level of
agreement between data and model calculations for the following four
asymmetries, as they involve chiral-odd GPDs~\cite{PrivCommPK}:
\begin{alignat}{2}
 &A_{\text{UT}}^{\sin(\phi-\phi_s)} \sigma_0 
 &=& -2 \im \Big[ 
     \epsilon \mathcal{M}^*_{0-,0+}\mathcal{M}_{0+,0+}
   + \mathcal{M}^*_{+-,++}\mathcal{M}_{++,++} 
   + \frac{1}{2} \mathcal{M}^*_{0-,++}\mathcal{M}_{0+,++} \Big], \label{eq:AUTphi-phiS}\\
 &A_{\text{UT}}^{\sin(\phi_s)} \sigma_0 
 &=& -\im \Big[ \mathcal{M}^*_{0-,++}\mathcal{M}_{0+,0+}
           - \mathcal{M}^*_{0+,++}\mathcal{M}_{0-,0+} \Big], \label{eq:AUTphiS}\\
 &A_{\text{UT}}^{\sin(2\phi - \phi_s)} \sigma_0  
 &=& - \im \Big[ \mathcal{M}^*_{0+,++}\mathcal{M}_{0-,0+} \Big], \label{eq:AUT2phi-phiS}\\
 &A_{\text{LT}}^{\cos(\phi_s)} \sigma_0
 &=&  - \re \Big[ \mathcal{M}^*_{0-,++}\mathcal{M}_{0+,0+} - \mathcal{M}^*_{0+,++}\mathcal{M}_{0-,0+} \Big] \, .  \label{eq:ALTphiS}
\end{alignat}
Here, the dominant $\gamma^*_L \rightarrow \rho^0_L$ transitions are described
by helicity amplitudes $\mathcal{M}_{0+,0+}$ and $\mathcal{M}_{0-,0+}$, which
are related to chiral-even GPDs $H$ and $E$, respectively. The subscripts $L$
and $T$ denote the photon and meson helicities $0$ and $\pm1$,
respectively. These GPDs are used since several years to describe DVCS and HEMP
data. The suppressed $\gamma^*_T \rightarrow \rho^0_T$ transitions are described
by the helicity amplitudes $\mathcal{M}_{++,++}$ and $\mathcal{M}_{+-,++}$,
which are likewise related to $H$ and $E$. By the recent inclusion of
transverse, i.e. chiral-odd GPDs, it became possible to also describe
$\gamma^*_T \rightarrow \rho^0_L$ transitions. In their description appear the
amplitudes $\mathcal{M}_{0-,++}$
related to chiral-odd GPDs ${H}_T$ ~\cite{Goloskokov:2011rd,PrivCommPK} and
$\mathcal{M}_{0+,++}$ related to chiral-odd GPDs $\xoverline{E}_T$
~\cite{Goloskokov:2009ia}. The double-flip amplitude $\mathcal{M}_{0-,-+}$ is
neglected. The transitions $\gamma^*_L\to\rho^0_T$ and
$\gamma^*_T\to\rho^0_{-T}$ are known to be suppressed and hence neglected in the
model calculations.

All measured asymmetries agree well with the calculations of
Ref.~\cite{PrivCommPK}. In Eq.~(\ref{eq:AUTphi-phiS}), the first two terms
represent each a combination of chiral-even GPDs $H$ and $E$. The inclusion of
chiral-odd GPDs by the third term has negligible impact on the behaviour of
\AUT, as can be seen when comparing calculations of
Refs.~\cite{Goloskokov:2008ib} and~\cite{PrivCommPK}. The asymmetry \AUT\ itself
may still be of small magnitude, because for GPDs $E$ in $\rho^0$ production the
valence quark contribution is expected to be not large. This is interpreted as a
cancellation due to different signs and comparable magnitudes of GPDs $E^u$ and
$E^d$ \cite{Adolph:2012ht}. Furthermore, the small gluon and sea contributions
evaluated in the model of Ref.~\cite{Goloskokov:2008ib} cancel here to a large
extent. The asymmetries \AUTphiS\ and \ALTphiS\ represent imaginary and real
part, respectively, of the same difference of two products
$\mathcal{M}^*\mathcal{M}$ of two helicity amplitudes, where the first term of
this difference represents a combination of GPDs $H_T$ and $H$, and the second a
combination of $\xoverline{E}_T$ and $E$. As can be seen in
Fig.~\ref{fig:asym_mean} and Table~\ref{tab:asym_summary}, while no conclusion
can be drawn on \ALTphiS\ because of larger experimental uncertainties, a
non-vanishing value for \AUTphiS\ is measured. The asymmetry \AUTphi\ represents
the same combination of GPDs $\xoverline{E}_T$ and $E$ as the second term in
\AUTphiS. The observation of a vanishing value for \AUTphi\ implies that the
non-vanishing value of \AUTphiS\ constitutes the first experimental evidence
from hard exclusive $\rho^0$ leptoproduction for the existence of transverse
GPDs $H_T$.

\section{Summary}
\label{sect:summary}
Asymmetries related to transverse target polarisation were measured in azimuthal
modulations of the cross section at COMPASS in exclusive $\rho^0$ muoproduction
on protons. The amplitudes of five single-spin asymmetries for unpolarised beam
and three double-spin asymmetries for longitudinally polarised beam were
extracted over the entire COMPASS kinematic domain as a function of $Q^2$,
$x_{Bj}$, or $p_T^2$. The asymmetry \AUTphiS\ was found to be $-0.019 \pm
0.008(stat.) \pm 0.003(syst.)$. All other asymmetries were also found to be of
small magnitude but consistent with zero within experimental uncertainties. Very
recent model calculations agree well with the present results. The results
represent first experimental evidence from hard exclusive $\rho^0$
leptoproduction for the existence of non-vanishing transverse GPDs $H_T$.

\section*{Acknowledgements}
We gratefully acknowledge the support of the CERN management and staff and the
skill and effort of the technicians of our collaborating institutes.  This work
was made possible by the financial support of our funding agencies.  Special
thanks go to P.~Kroll and S.~Goloskokov for providing us with the full set of
model calculations as well as for the fruitful collaboration and many
discussions on the interpretation of the results.

\end{document}